\documentclass[sigconf,10pt]{acmart}

\usepackage[english]{babel}
\usepackage{xcolor}
\usepackage{amsfonts}
\PassOptionsToPackage{hyphens}{url}
\usepackage{hyperref}
\usepackage{color}
\usepackage{graphics}
\usepackage{graphicx}
\usepackage{url}
\usepackage{listings}
\usepackage{multicol}
\usepackage{multirow}
\usepackage[scaled]{helvet}
\usepackage{rotating}
\usepackage{xspace}
\usepackage{comment}
\usepackage{enumitem}
\usepackage{amsmath}
\usepackage{mathrsfs}
\usepackage{cancel}
\usepackage{cleveref}
\usepackage{algpseudocode}
\usepackage{algorithm}

\newcommand{\sysName}{Spyglass\xspace}

\newcommand{\edalgo}{Searchlite\xspace}
\newcommand{\spalgo}{SSFP\xspace}
\newcommand{\spalgofull}{switch-synchronous Fourier processing\xspace}

\crefformat{section}{\S#2#1#3} %
\crefformat{subsection}{\S#2#1#3}
\crefformat{subsubsection}{\S#2#1#3}

\usepackage{subcaption}

\usepackage[explicit,noindentafter]{titlesec}
\titlespacing\section{4pt}{4pt plus 2pt minus 2pt}{1pt plus 2pt minus 2pt}
\titlespacing\subsection{4pt}{4pt plus 2pt minus 2pt}{1pt plus 2pt minus 2pt}
\titlespacing\subsubsection{4pt}{4pt plus 2pt minus 2pt}{1pt plus 2pt minus 2pt}

\acmDOI{}
\acmISBN{}
\acmConference[arXiv Preprint]{}
\acmYear{2026}
\copyrightyear{}
\acmPrice{}

\renewcommand\footnotetextcopyrightpermission[1]{} %
\setcopyright{none}

\title{\sysName: Directional Spectrum Sensing with Single-shot AoA Estimation and Virtual Arrays}
    
\author{Raghav Subbaraman, Akshit Agarwal, Wenhao Chen, Dinesh Bharadia}
\affiliation{
    \institution{University of California San Diego}
}
\email{{rsubbaraman, akagarwa, wec035, dineshb}@ucsd.edu}
\renewcommand{\shortauthors}{R. Subbaraman, A. Agarwal, W. Chen, D. Bharadia}
\date{Feb 2026}

\begin{abstract}
In this paper, we introduce \sysName, a spectrum sensor designed to address the challenges of effective spectrum usage in dense wireless environments. \sysName is capable of observing a frequency band and accurately estimating the Angle of Arrival (AoA) of any signal during a single transmission. This includes additional signal context such as center frequency, bandwidth, and I/Q samples. We overcome challenges such as the clutter of fleeting transmissions in common bands, the high cost of array processing for AoA estimation, and the difficulty of detecting and estimating channels for unknown signals. Our first contribution is the development of \edalgo, a protocol-agnostic signal detection and separation algorithm. We use a switched array to reduce cost and processing complexity, and we develop \spalgo, a signal processing technique using Fourier transforms that is synchronized to switching boundaries. \sysName performs multi-channel blind AoA estimation synchronized with the array. Implemented using commercially available hardware, \sysName demonstrates a median AoA accuracy of 1.4$^\circ$ and the ability to separate simultaneous signals from multiple devices in an unconstrained RF environment, providing valuable tools for large-scale RF data collection and analysis.
\end{abstract}

\begin{document}

\maketitle

\textbf{}\section{Introduction}
\label{introduction}

The growing number and accessibility to wireless devices have opened up a variety of privacy and security implications. Light-bulbs and wall-clocks, seemingly innocuous devices in our homes are connected wirelessly to the world's internet. When temporarily occupying an unfamiliar area like a hotel room or office, we need to worry about hidden wireless cameras or microphones that put private information and our safety in jeopardy~\cite{nythidden}. Regulating or identifying unauthorized wireless transmissions in sensitive areas like top-secret facilities or access-controlled buildings is increasingly difficult as \textit{any} electronic device could potentially be a threat. For less than \$300 purchase in the open market, an attacker can gain unauthorized access to wireless RFID systems and even wirelessly crash iPhones~\cite{flipperzerobad}. In the effort to identify potentially harmful wireless transmissions and devices, a tool that can "see" and distinguish these transmissions is a very useful. Think of this as a wireless camera of sorts, one that allows the end user to sift through the activity in an area and hone in on something that seems suspicious. 

To build such a "wireless camera" tool, it is important to understand how devices use radio frequency (RF) spectrum in time, frequency, and space. Wireless devices transmit energy in RF spectrum, usually around a specific center frequency and with a limited bandwidth. The exact contents of this energy, how it is modulated or demodulated is usually specified in a (sometimes proprietary) specification or protocol. Devices duty-cycle their transmissions in accordance to data requirements, and may switch off to save power. The RF spectrum consists of a cacophony of all these devices transmitting at different frequencies and time. Therefore, the first requirement for our tool is to \textit{understand how RF spectrum is used in time and frequency, without any prior assumption on which device is transmitting or what protocol it is using} (R1). The wireless devices we wish to see are also distributed in space within our region of observation. A hidden camera could use WiFi transmissions, the same as our computer; but the fact that it originates from an unauthorized location makes it dangerous. Knowing the rough origin of a suspicious transmission makes it easier weed it out. Thus, the second requirement for our tool is to \textit{understand the spatial context or direction of RF transmitters (R2)}. Finally - similar to a CCTV camera - \textit{our tool should be sufficiently low-cost and easy to produce so that it can be deployed widely (R3)}.

 \begin{figure}[!t]
    \includegraphics[width=\linewidth]{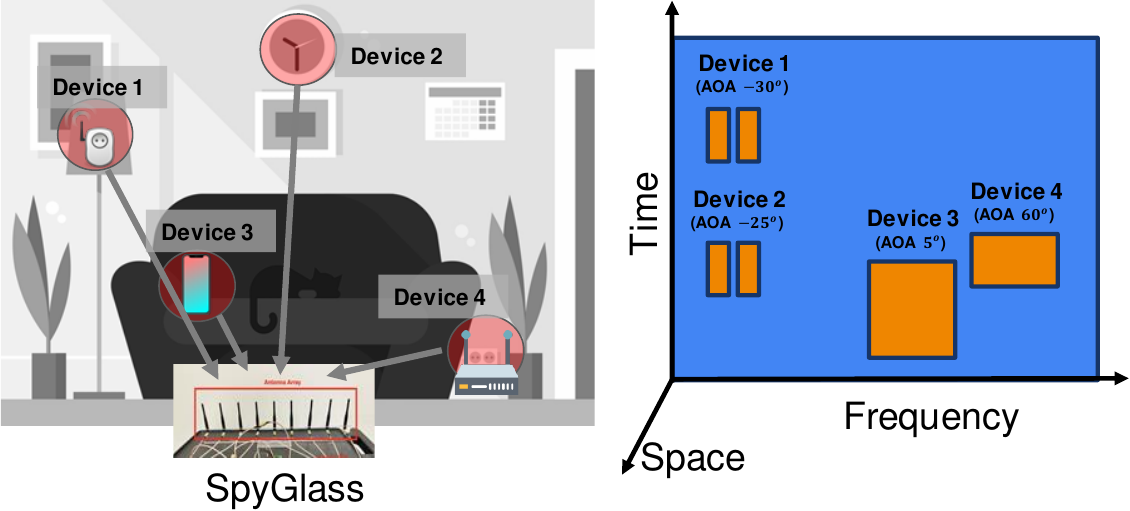}
    \caption{\small{\sysName can estimate the angular locations of any wireless transmitter in the environment. It uses time-frequency energy detection, synchronized DSP and switching arrays to achieve its function in real-time with efficient hardware use.}}
    \label{fig:figure1}
\end{figure}

State-of-the art RF spectrum sensors, localization systems, and commercial direction finders fail to meet all of the requirements that we require for the "wireless camera" tool. In the last decade, new spectrum sensing platforms have been developed, mostly focusing on wide-bandwidth and real-time observation at low-cost~\cite{ss1,bigband2,sparsdr}. While these platforms provide RF spectrum data in the form of I/Q samples or power spectral density (PSD), they do not provide methods to separate spectrum activity nor allow direction finding; thus failing to meet R1 and R2. Commercial signal direction finders are used for specialized applications such as interference hunting or for defense purposes~\cite{rohdeschwarzdirection}. These devices are made for extremely rugged use, are expensive, and require prior knowledge of the transmitter or protocol to work. Signal direction finders do not meet R1 and R3. The use of multi-antenna arrays, especially switched antenna arrays, is the approach of choice for low-cost angle of arrival (AoA) estimation and localization systems~\cite{tyrloc,ubicompbleswitch}. These systems meet R2 and R3, but do so by making assumptions specific to a combination of protocols like  WiFi~\cite{wifiswitching-mobicom}, BLE~\cite{ubicompbleswitch}, or LoRa~\cite{tyrloc}. These techniques do not generalize to unknown protocols, and therefore do not meet R1.

In this work, we introduce \sysName (Fig~\ref{fig:figure1}), the "wireless camera" tool that achieves meets requirements R1-R3 by leveraging switched antenna arrays for spectrum sensing, and developing specialized signal processing tools for protocol agnostic AoA estimation despite switching. To achieve protocol agnosticism, we develop \edalgo, an algorithm that automatically detects and separates signals in the time-frequency plane by performing energy detection on a short-term-fourier-transform (STFT) spectrogram (meets R1). We also develop \spalgofull (\spalgo), a signal processing framework to perform reversible STFT aligned with switching boundaries. \spalgo and \edalgo put together allow us to separate transmissions and perform individual signal processing. By leveraging the guarantees provided by \spalgo, we can estimate the relative phases on our antenna array, and therefore the AoA of each transmission (meets R2). \sysName uses a switched antenna array in order to create large arrays (i.e. better AoA estimation capability) at lower cost (meets R3). Our prototype for \sysName is open and modular, including software implementations of \edalgo and \spalgo, and evaluated on real-world devices.

\textbf{\edalgo:} In order to deliver on R1, we need to somehow capture and separate transmissions from different protocols before we do any sort of processing on them. This separation needs to happen in the time-frequency domain, and could provide additional context of the possible contents/protocol of the signal. A simple method to do this is to take stock of all the frequency channels of known protocols. We can then use filters, resamplers and matched filters to detect transmissions of any of these protocols~\cite{revathy_cloudsdr}. This approach is difficult due to the processing overhead of running an arbitrary number signal processing that scales with the number of known protocols. Further, it works only for protocols that we already know of, and will not identify unknown transmitters. On the contrary, we need a method that automatically cover all possibilities without assumptions on protocol. \edalgo is an automatic, wideband energy detection and separation algorithm. Inspired by SparSDR~\cite{sparsdr}, we use a reversible STFT spectrogram as the representation for energy detection. We develop a noise estimation model, and use 2-D object detection to separate transmissions. \edalgo is implemented in Python for real-time throughput and works stand-alone with multi-antenna I/Q streams.

\textbf{\spalgo}: After detecting and separating transmissions in the STFT domain using \edalgo, we need to extract the each of them for further individual processing. Performing arbitrary extraction of sub-arrays from the STFT requires careful inversion of the transform and phase compensation~\cite{sparsdr}. This inversion is further complicated by the requirement to use a switched antenna array. In order to correctly perform the STFT inversion while switching, we would need to carefully map switching events to the STFT matrix. A straightforward way to design this would be to switch slowly and perform STFT between each switch boundary - avoiding the problem altogether. Typical spectrum sensing applications require at least 1024 bin STFTs~\cite{sparsdr} due to frequency resolution and processing efficiency reasons. At wideband processing rates of 30 MHz, this corresponds to a $\approx$30 us time window. Transmissions from common protocols like WiFi and BLE are typically only 100-200 us~\cite{crescendo} long, making it impossible to switch more than 3-6 antennas within the length of the packet. We therefore \textit{have to switch antennas at rates faster than a single FFT size}. \spalgo provides a DSP framework for the choice for switch rate, FFT size, and inverse FFT size for a switching stream. \spalgo guarantees that the time domain representation of a sub-array of the STFT \textit{will have integer number of samples for every switch event and that switch boundaries happen on sample-boundaries}.

\textbf{AoA estimation and Prototype:} To estimate the AoA from measurements from our switched antenna, we need to estimate the relative phase between the antennas for the \textbf{same} signal~\cite{arraytrack}. Since we switch, we never get the same signal on each antenna. Therefore, we use a reference antenna that samples simultaneously (without switching). Using the reference antenna, we can get pair-wise phase difference for each antenna and then compute AoA. We use a 2 Rx radio with synchronized receive chains to achieve this. One receiver is used as the reference, and the other one switches. \sysName is implemented using COTS hardware: a USRP B210 SDR~\cite{usrp}, an RF switch, and a synchronizing FPGA. \edalgo is written in Python, runs in real-time and can be used stand-alone with any software defined radio (SDR). Our implementation of \sysName is modular, and we intend to release the software for \edalgo and the hardware implementations in public domain upon acceptance of this work. We believe \sysName, and especially \edalgo are valuable tools to collect, verify, and label RF data on a large scale. We intend to open source the source code for \edalgo upon publication of this work, for access please contact the authors.

We evaluate \sysName in an unconstrained indoor environment with multiple devices:
\begin{enumerate}
    \item \sysName provides a 90th percentile AoA accuracy of 3.3$^\circ$, and median accuracy of $1.4^\circ$, 1.7x better than baseline 
    \item \sysName can estimate 8-antenna AoA for packets as small as 50 us. For smaller packets, performance degrades gracefully as it is equivalent to AoA estimation with lesser antennas.
    \item \sysName can separate devices operating simultaneously in a band and provide frequency, time context in addition to AoA.
\end{enumerate}

\section{Background}

\begin{figure}[t!]
\centering
\includegraphics[width=.5\textwidth]{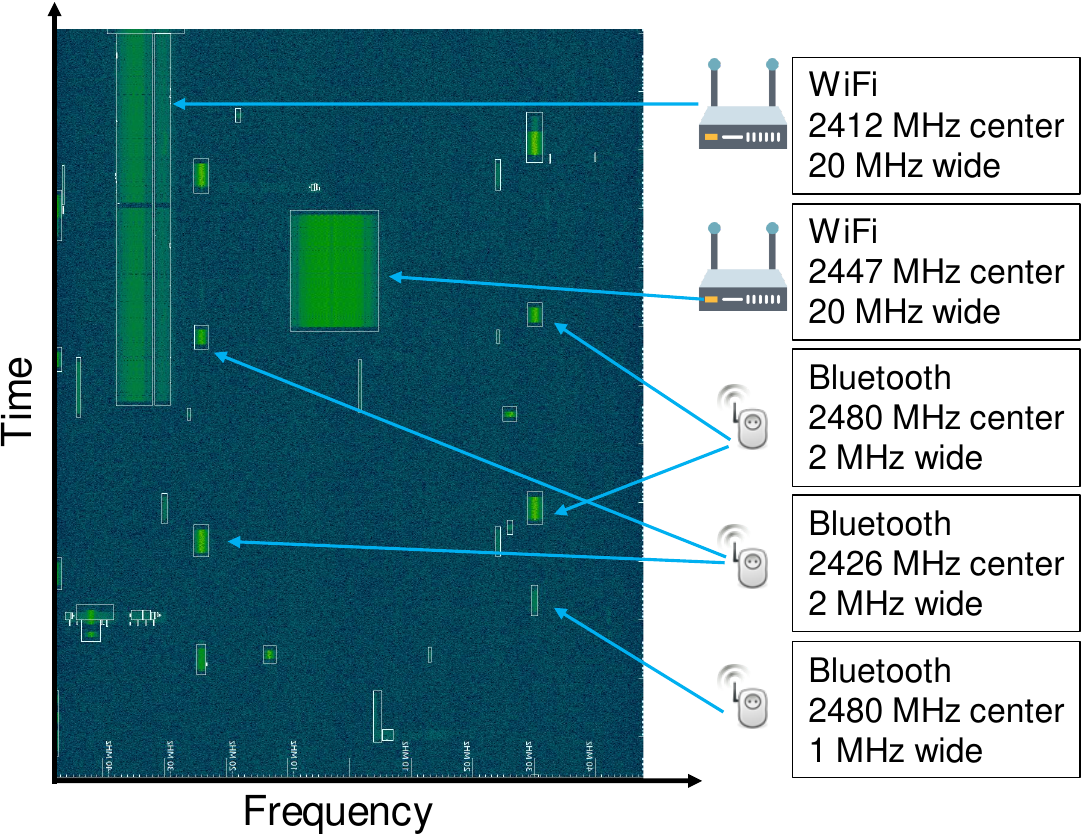}
\caption{\small{100 MHz RF spectrum centered at 2450 MHz captured with an SDR visualized using a spectrogram~\cite{inspectrum}. Regions of brighter color correspond to time-frequency instants where energy is high. Different devices operate using different protocols and share the spectrum. The white annotations on the image are automatically performed using \edalgo.}}
\label{fig:background}
\end{figure}

The RF spectrum is a busy home to a variety of chirpy devices. Commonly used bands like 2.4 GHz ISM are extremely crowded with tens of devices multiplexed within sensing distance in typical indoor scenarios. To support large data-rates, and to avoid collisions, many signals use short packets of the order of few 100 us, and sometimes even hop around different frequency bands. Real-time analysis of a portion of the RF spectrum can be performed using a relatively wide-band (10s of MHz) radio that provides complex baseband I/Q samples. Such configurable software defined radios (SDRs) have become affordable and commonplace in the last decade~\cite{limesdr,antsdr,plutoplus,usrp}. For \$350, one can get a 2x2 channel SDR with 61.44 MHz of sampling bandwidth~\cite{antsdr-cheap350}.

When analyzing a large and complex portion of the RF spectrum, spectrum analysis methods need to be used. Spectrum analysis methods provide a way to process the I/Q samples provided by the SDR and understand the contents of the RF spectrum as seen by the radio. These tools are typically based on transforms like the FFT or polyphase filter banks, and can separate activity into frequency "bins"~\cite{fredSTFT}. Performing spectrum analysis in short windows over the I/Q sample stream can allow us to visualize the time-evolving nature of spectrum use as show in in Figure~\ref{fig:background}. 

Every transmitting device emits RF energy, whose frequency characteristics are defined by the signal type of the device and time characteristics depends on when the device transmits. For example: WiFi devices that operate in 2.4 GHz typically operate in one of three different bands in the US (CH1: 2412 MHz, CH6: 2437 MHz, CH11: 2462 MHz), and their energy is localized to a bandwidth of 20 MHz around their respective bands. BLE advertisement packets used for discovery, contact tracing, etc are 2 MHz wide and centered around one of three bands (2402 MHz, 2426 MHz, 2480 MHz). Spectrum analysis as in Figure~\ref{fig:background} can be used to inform when such devices are transmitting, allowing an expert to go in and filter out the signal of interest for further processing. 

Naturally, there is a lot of value in automating spectral analysis. An algorithm that can observe wideband spectrum and separate signals with little-to no prior information is useful for flexible compute and processing. There has been a lot of interest in fields like automatic modulation recognition, signal separation in the presence of interference, and cyclostationary separation~\cite{mf-cyclo-compare}, these works focus on a specific problem of interference or channel, not generic-unconstrained spectrum analysis. To the best of our knowledge, there is no publicly accessible tool that automates signal separation like \edalgo can in Figure~\ref{fig:background}.

\subsection{Beyond spectral analysis: AoA}

Automating spectral analysis provides time and frequency context such as center frequency, bandwidth, and packet size. However, this is not enough to fully understand spectrum use. Different devices that operate in the same band emit packets that look the same albeit at different times, making it easy to mistake one for the other. Understanding additional context like the Angle of Arrival (AoA) is necessary to separate devices and enable applications that rely on position like localization and interference modeling for spectrum sharing~\cite{rf_propa_model}. 

Allowing AoA computation requires extending spectral analysis to multiple simultaneous streams from antennas. As discussed in the previous section, multi-antenna setups quickly become intractable due to cost of hardware and processing. Single antenna SDRs are available for as low as \$ 250~\cite{plutosdr}, while two antenna SDRs cost \$ 350~\cite{antsdr-cheap350}. Four antenna SDRs that are used for educational and research purposes are at \$ 10k+~\cite{urspn310}, and require two 10G network interfaces to sustain at full-rate!

Looking at Figure~\ref{fig:background}, we see that packets from each device exist for at least few 10s of us. Our goal was to try and reconfigure our hardware to emulate a large antenna array by changing its behavior during the packet duration. Prior systems have experimented with switched antenna arrays for specific protocols like WiFi, Bluetooth, and LoRa~\cite{ubicompbleswitch,tyrloc,wifiswitching-mobicom}. In \sysName, we achieve protocol agnostic sub-packet AoA by leveraging fast RF switching synchronized with the baseband.

\begin{figure*}[ht]
\centering
\includegraphics[width=\textwidth]{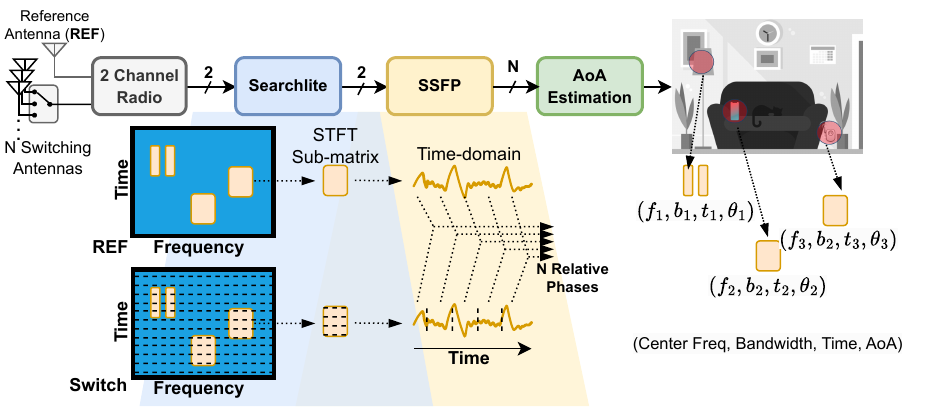}
\caption{End-to-end architecture of the \sysName system, showcasing the signal processing pipeline from input to output. The diagram illustrates the integration of multi-antenna detection, switched antenna array, and angle-of-arrival measurements. Each component and interface is labeled to highlight the data flow and processing stages, providing a comprehensive view of the system's functionality. Below the block diagram, a view of the time-frequency plane is shown, where energy detection, and subsequent relative channel estimation is highlighted.}
\label{fig:design}
\end{figure*}
\section{Design}

\sysName's goal is to separate signals in the environment, then, provide time, frequency and spatial context. While it is possible to estimate time-frequency properties of a signal using data from a single receiver, understanding spatial context in the form of AoA requires processing multiple high-rate streams from numerous antennas and receivers. To keep const and complexity low, we simplify this architecture using a switch to simulate a virtual antenna array with just one receiver. While switched arrays have been used before for AoA estimation, it is challenging to extend this function in an environment with multiple simultaneous, unknown transmitters. We developed \edalgo and \spalgo to solve these challenges.

In, \edalgo, we leverage an STFT to represent the time domain I/Q stream into a 2-D spectrogram where signals can be separated. We then use an automatic energy detection algorithm to detect regions of activity within the spectrogram and annotate the sub-matrices. \edalgo is designed to synchronously operate on multiple simultaneous stream, allowing it to be used easily with multi-antenna systems. \edalgo's annotated sub-matrix on the reference stream can be readily inverted using an ISTFT to produce a time-domain I/Q stream for further processing/decoding. However, \edalgo's annotated sub-matrices on the switched spectrogram contains embedded switching events. These events need to be identified and mapped to the source antenna so that a virtual array can be created. We develop the \spalgo framework to demonstrate the need of synchronization and careful selection of switch-times and FFT sizes that guarantee predictable switching events. In addition, we show how partial ISTFTs can be performed on the sub-matrices while maintaining the predictability guarantee. Partial ISTFTs significantly reduce the amount of processing load on the system by simply reducing the FFT size to the minimum required to re-construct time-domain I/Q stream while satisfying the Nyquist criterion. \spalgo finally provides time-domain I/Q for the reference and switched streams for each signal. We measure the relative phase between the switched I/Q stream and the reference I/Q stream to simulate a virtual array and estimate AoA for every signal. The end-to-end design flow of \sysName is presented in Figure~\ref{fig:design}.

\subsection{Signal agnostic detection and separation: \edalgo}

\begin{figure}[t]
\includegraphics[width=\linewidth]{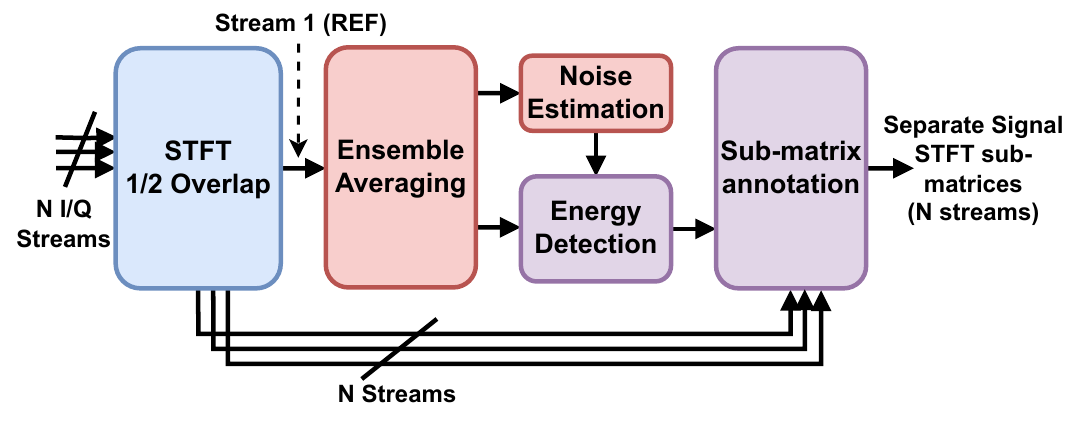}
\caption{Architecture of the \textit{\edalgo} algorithm illustrating the signal processing pipeline. The process begins with performing STFT on each input signal channel, followed by ensemble averaging, noise estimation, and energy detection on the reference channel. Ultimately, specific areas of energy are extracted synchronously on all channels, providing multi-channel signal agnostic detection.}
\label{fig:edalgo}
\end{figure}

\edalgo performs multi-channel protocol agnostic signal separation using energy detection in the time-frequency plane. It uses the first stream of IQ (from antenna 1) as reference to perform energy detection. The data from all the streams go through the processing pipeline illustrated in Figure~\ref{fig:edalgo}:

\noindent
\textbf{STFT}: \edalgo uses a half-overlap windowed STFT that uses a Constant-Overlap-Add (COLA) window~\cite{fredSTFT, welchstft}. 
The output of the STFT can be interpreted as a 2-D matrix with time and frequency as each of the dimensions. The frequency axis is referred to as "bins". STFTs with this property are fully invertible, i.e, we can retrieve the respective time-domain I/Q of any sub-band of the STFT by performing an inverse-STFT on a contiguous rectangular region in the 2-D matrix. We use the Hann window in our STFT~\cite{fredSTFT}. For brevity, we omit other analysis and refer the reader to SparSDR~\cite{sparsdr}, whose authors detail why this STFT is fully invertible and discuss the choice of windowing functions.

\noindent
\textbf{Ensemble averaging}: The output of the STFT is a complex number, and to detect energy, we convert it into power spectral density (PSD - a measure of energy) by taking the square of the absolute value. In this domain, the variance of each element in the output 2-D matrix of the STFT is very large~\cite{fredSTFT}. Ensemble averaging (averaging across time-domain) is necessary to reduce the variance of the PSD estimates. This operation ensures that both signal and noise power spectral densities have lower variance - improving the ability to separate signals from noise. The ensemble averaged output can be used to perform noise floor estimation.

\noindent
\textbf{Noise floor estimation}: The next step is to estimate the power of the noise floor so that we can separate signals from noise. This is not straightforward as the STFT output contains both signals and noise: reduction operations like mean and median will be affected by the signals. We instead opt to use the minimum value of the PSD on a per-frequency bin basis as the reference point for noise floor estimation. Signals in various bands turn on-and-off, so it is likely that there will at least be a few instances of each bin containing only noise, and the minimum operation will find it. In bands where there are signals present always, the minimum value is still not good enough to estimate the noise floor. For these bands, we use the fact that the noise floor is typically flat across few 10s of MHz. We identify outliers in the min vector and perform interpolation across known good frequency bins to get a good noise floor estimate even for these bands. Now that we have the minimum value for each frequency bin, we use statistical properties of gaussian noise to estimate the mean noise power from the minimum value.

\noindent
\textbf{Energy detection}: Once we have the mean noise power estimates for each frequency bin, we perform a simple threshold to mark all bins that are above the threshold as containing energy. This operation produces a bit-mask on the 2-D matrix output of the STFT. We then use binary morphological operations to remove completely unconnected pixels and connect nearby active pixels that are not connected~\cite{szeliski2022computer}. Then we use traditional computer vision object finding to extract disconnected objects from the resultant bit-mask~\cite{scipy,szeliski2022computer}. The extracted disconnected objects are bounds of 2-D sub-matrices of the STFT output where a signal is present. This is how we separate each signal from the STFT from noise and from other signals.

\noindent
\textbf{Multi-antenna STFT sub-matrix extraction}: Now that we have labelled areas in the STFT matrix that contain energy, we excise these sub-matrices from each STFT. Excision of the exact same sub-matrices from each receive channel is important to maintain phase synchronization. 

By integrating these functions and interfaces, \edalgo offers a streamlined and efficient process for signal agnostic detection on multiple antennas. The modular design ensures easy integration with external systems, and the clear delineation of function makes it adaptable to different applications.

\subsection{\spalgo: recover switched time-domain for each signal}

In the previous section, we went over how distinct signals in the STFT matrix can be detected and represented as disjoint sub-matrices using \edalgo. Since we need to perform processing on time-domain I/Q to understand explicit switching boundaries, we need to invert the ISTFT for each disjoint submatrix. A straightforward inversion method would be to zero out everything except a sub-matrix of interest and perform an ISTFT, yielding time-domain I/Q at the full sample rate. 
In this section we describe \spalgofull, a framework to select Fourier transform parameters and perform partial ISTFTs that provides guarantees on switching boundaries on the inverted time-domain I/Q. 

\noindent
\textbf{Switch time and FFT size selection:} For simplicity of processing, we need to guarantee that switch boundaries are at sample boundaries and that we have an integer number of switch events within an inverse transform. Therefore, the switching time $T_{SW}$ (i.e dwell time of sampling on each antenna) and the FFT size need to be chosen carefully. In typical implementation (including ours), the switching logic will be clocked by the reference clock (ref-clk) of the radio, whose frequency is $F_{REFCLK}$ (typ. 10s of MHz). The sampling clock of the radio is generated using a PLL based on this ref-clk, and has frequency $F_{S}$. To make sure that switch boundaries align with sampling clock boundaries, the switch time should adhere to the following equation:

\begin{equation}
    T_{SW} = \frac{k}{gcd(F_{REFCLK},F_{S})}, k \in \mathbb{N}
\end{equation}

To ensure that the FFT boundaries have an integer number of switch cycles, it follows that the FFT size $NFFT$ is constrained by:

\begin{equation}
    NFFT = p \times T_{SW} \times F_{S}, p \in \mathbb{N}
\end{equation}

Larger values of $NFFT$ may need to be chosen in \edalgo for better frequency resolution.

\noindent
\textbf{Efficient partial sub-matrix inversion:} Performing a full ISTFT with $NFFT$ points is wasteful, as typical occupancy of signals within a wideband SDR's sampling bandwidth is less than 10\%~\cite{sparsdr}. Partial ISTFTs can be used to reduce the DSP compute load, while simultaneously recovering the baseband representation of signals~\cite{sparsdr}. It is difficult to perform partial-ISTFTs on a switched stream while guaranteeing that switch boundaries happen. The size of the sub-matrix to invert is arbitrary and signal dependent. Performing a naive sub-matrix inversion will lead to switching boundaries aligned with fractional samples and incomplete switch cycles in the time-domain waveform. 

To solve this challenge, we use the fact that sub-matrix ISTFTs can be padded arbitrarily with zeros. We observe that if the sample rate of the inverted sub-matrix multiplied by $T_{SW}$ is an integer, then our guarantees will be met. To enforce this constraint, we restrict the partial IFFT size ($NIFFT_{partial}$) as follows:

\begin{equation}
    NIFFT_{partial} = q\times p, q \in \mathbb{N}
\end{equation}

Therefore, $NIFFT_{partial}$ should be a multiple of $p$, the number of switch boundaries contained in $NFFT$ samples at full sample rate $F_{S}$.

\noindent
\textbf{Note on synchronization and group delay:} In order for \spalgo to work as intended, synchronization between the switching clock and the sampling clock is necessary. In addition, the STFT-ISTFT algorithm implementation should have an exactly predictable group delay. Unknown group delays in processing can mis-align switch boundaries from their expected location and destroy estimation accuracy. Our implementation of the half-overlap STFT and partial ISTFT ensures that the group delay is zero for the two when put together - ensuring that \spalgo's guarantee is maintained.

\subsection{AoA estimation}

Unique time-domain I/Q for the reference and switched streams can be extracted for each signal using \edalgo and \spalgo. But we are not done yet, in order to estimate the AoA of the signal, we need relative phases across the virtual antenna array. We never receive the same signal on each of the antennas due to the construction of the switched array, each contiguous part of the switched I/Q stream is from a different antenna port. We leverage the reference stream to compute the relative phase between the reference and each of the switching antennas. Since the reference stream contains an exact, synchronized copy of samples for each switch-port, the phase difference can be computed by taking a conjugate array product between the samples of the two streams.

The AoA computation block receives the samples and metadata from the \spalgo algorithm. AoA estimation using the IQ samples was done using the Maximum Likelihood Estimation (MLE) algorithm. For a pair of I/Q time series, the following algorithm is followed to compute the AoA. The algorithm uses the 2 streams of ($IQ_{REF}$ and $IQ_{SW}$), the center frequency of the energy detected ($f_c$) and the fixed antenna spacing vector (d) to compute AoA. For each contiguous transmission, the algorithm gives one estimate of AoA per full-cycle of N switches. 
\begin{algorithm}
\caption{AoA Computation through MLE}
\begin{algorithmic}[1]

\State $Phase = IQ_{REF}*IQ_{SW}$ 
\State $\phi = Vector \ with \ Relative \ Phases$
\State $\phi = [\phi_{ant1} \ \ \phi_{ant2} \ \  ... \ \   \phi_{antN}]$
\State $Define \ \ \alpha = \frac{2\pi d sin(\theta_{estimates})}{\lambda}$ \Comment{$\theta_{estimates} = [-\pi/2, \pi/2]$}

\State $T = e^{j \alpha n}$
\State $DFT = T \times e^{i\phi*}$
\State $AoA = \arg \max_{\theta_{estimates}}(|DFT|)$ 

\end{algorithmic}
\end{algorithm}

\section{Building \sysName}

\subsection{Hardware Design}
\begin{figure}[h!t]
    \includegraphics[width=\linewidth]{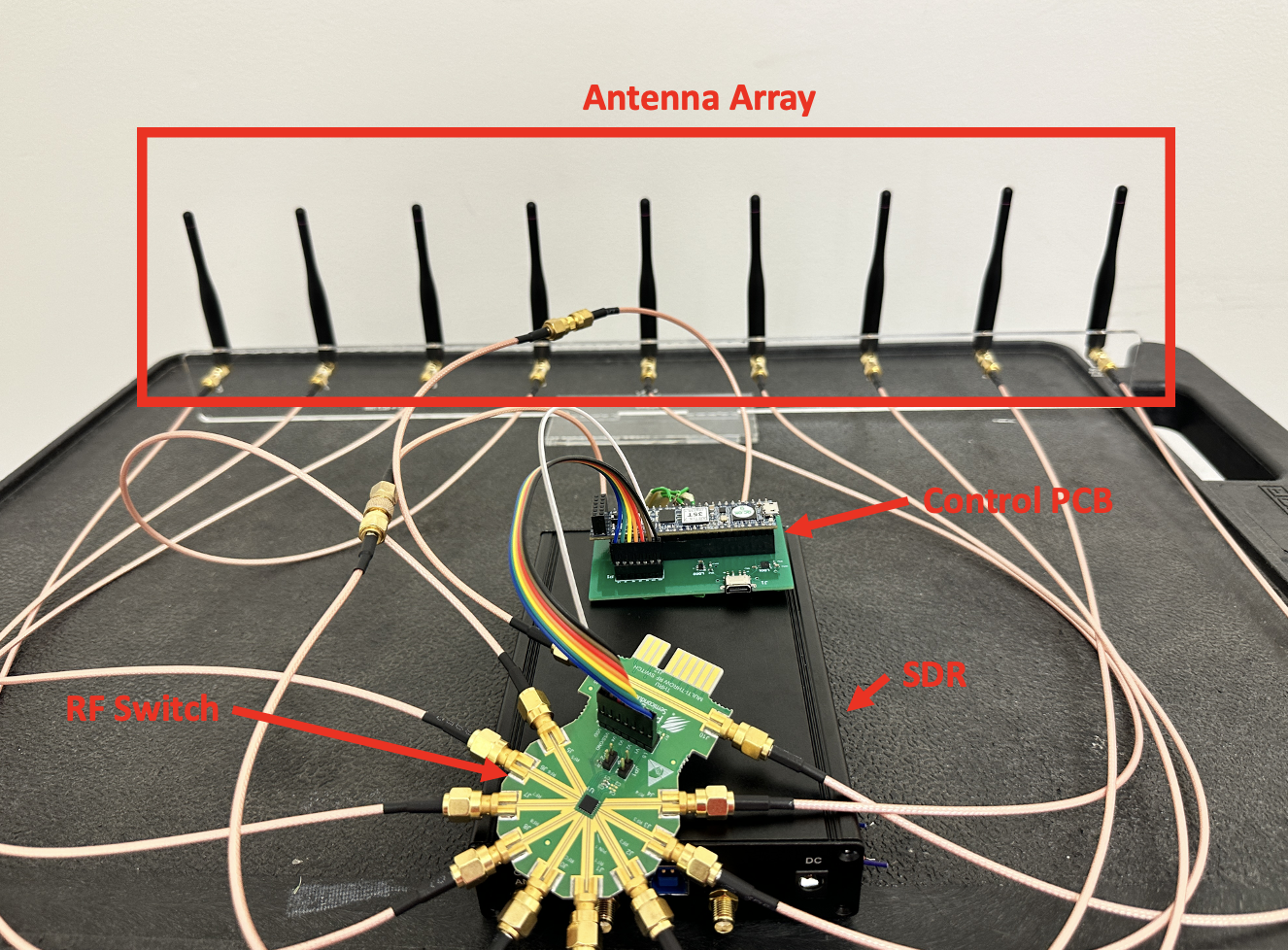}
    \caption{\small{Full \sysName hardware system featuring the switched antenna array, SDR, and the control PCB}}
    \label{fig:hardware}
\end{figure}

\noindent
\textbf{Switched Antenna Array -} The switched antenna array was implemented using standard COTS antennas and the PE42582 RF switch evaluation board \cite{PE42582, EK42582-02}. This switch is an SP8T switch and has an operational frequency from 9kHz to 8GHz. The switch is also capable of fast switching times that is essential for switching multiple times during the course of a single packet and control logic can be provided externally to toggle between ports.The 8 RF input ports are connected to the antennas and the RF common port is connected to one of the receive chains on a USRP B210 SDR \cite{usrp}. The antennas are mounted on a custom laser cut acrylic plate. This plate can be modified to have any geometry with up to 8 antennas in any configuration, including 2-D geometries for multi-axis AoA. One end of the calibration trace on the evaluation board is connected to an additional antenna (reference antenna) and the other end of the trace is connected to the second receive chain on the B210. This stream acts as the reference IQ data for signal identification. A picture of the entire setup is shown in Fig~\ref{fig:hardware}.

\noindent
\textbf{Control PCB -} A custom PCB was used to control the port connected to the common port on the RF switch evaluation board and provide necessary bias voltages for the switch. LDOs \cite{3v4_LDO}\cite{minus3_LDO} were used to generate a 3.4V VDD for the switch and a -3V boas voltage for fast switching operation of the particular model of switch used. A CMOD A7 35t FPGA \cite{cmod} is used on the PCB for port selection for the RF switch. The timed switching is implemented using a Finite-State Machine on the FPGA which is controlled by a 40MHz reference clock, a trigger and a reset signal from the B210 SDR.

\noindent
\textbf{Control Logic - } While the reset signal is at logic high, a control signal is generated such that none of the 8 input ports on the switch is connected to the common port of the switch. At the logic low of the reset, the FSM waits for the trigger pulse to start switching. At the trigger pulse, the FSM starts the switching cycle by generating the required 3-bit control logic signals V1-V3 for the current antenna. V1, V2 and V3 control the input port that is connected to the common port. Switching happens periodically according to a parameter switchtime which is sent to the FPGA using a UART interface between the FPGA and the python interface before the trigger signal is sent. Every switchtime seconds, the 4-bit control signal is incremented by 1 which corresponds to a switch from the current antenna to the next. At any point, if the reset signal is pushed to a logic high, the system stops switching and goes back to the initial state.   

\subsection{Synchronization}
The system design is very contingent on the synchronization of different pieces of hardware to know which IQ sample in the switched data stream corresponds to which antenna on the switched array. To make this happen, it was important to synchronize all hardware components with individual clocks, namely the laptop, the SDR, and the FPGA. Precise UHD commands timed commands were used to control the B210 and to acquire IQ samples.   

\begin{figure}[h!t]
    \includegraphics[width=\linewidth]{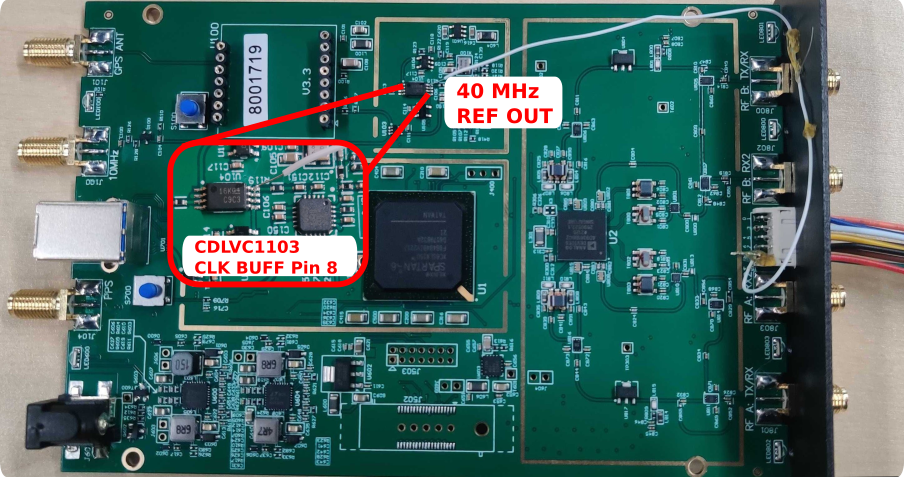}
    \caption{\small{An illustration of the modification performed on the USRP B210 to allow the REF Clock to be exposed for external FPGA synchronization.}}
    \label{fig:b210_mod}
\end{figure}

\begin{figure}[h!t]
    \includegraphics[width=\linewidth]{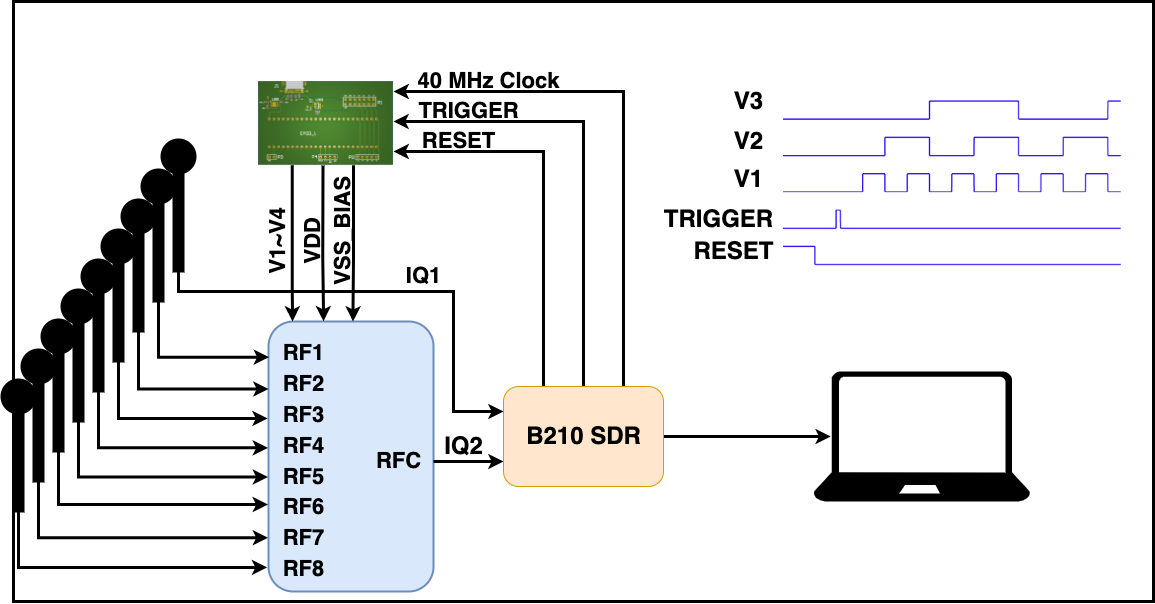}
    \caption{\small{Synchronization diagram of \sysName. The USRP B210 sends a trigger to the control PCB at the same time when the ADCs start sampling. The control PCB switches the antenna array by controlling the state of the RF switch. The control PCB is synchronized to the sample clock of the USRP by sharing its reference}}
    \label{fig:implfig}
\end{figure}

The GPIO pins on the B210 were used to synchronize the SDR and the FSM on the CMOD FPGA. Three signals, namely a 40 MHz reference clock, a trigger signal and a reset signal were used. Since there is no reference clock on the GPIO pins of the B210, one of the grounded IO pins was desoldered and was soldered to the output of an unused clock buffer on the B210's PCB (Fig~\ref{fig:b210_mod}). With a reference clock available, timed UHD commands were used to generate reset signal for the FSM and trigger signal indicating the start of the switching cycle (Fig~\ref{fig:implfig}). 

When the system is started, all required parameters are sent to the devices through the interfaces mentioned earlier. The reset is then set to low and the trigger is sent after a small delay. As soon as the trigger is detected in the FSM, the switching begins where the FPGA produces the relevant control signals according to the antenna that is supposed to be connected.

\subsection{Host-Software}
The host software for \sysName is divided into the 3 major parts: The first part is the interface from SDR to the laptop. GNURadio along with a ZMQ pipeline \cite{ZMQ} is used to transfer data from the SDR. In every cycle, the interface reads a specified number of samples from the buffer of the SDR and sends them to the energy detection algorithm for signal identification. This interface also implements the reset and trigger signals used by the FPGA before starting collecting samples. 

The second part is the energy detection algorithm itself. Implemented in python, it takes in chunks of N samples in each cycle and outputs the IQ samples associated with detected energies through a separate ZMQ interface. The STFT, ISTFT ensemble averaging, noise estimation, and energy detection are performed by composing functions from Python's \texttt{scipy} and \texttt{numpy}~\cite{scipy, numpy} libraries.

The last part is the AoA Computation block. This block gets IQ data of detected energies from the previous block. It uses the IQ samples from the 2 input channels to compute the AoA for the given signal. The IQ samples, the metadata for the signal and the AoA computations can be stored for future analysis or even decoding.

\section{Evaluation}

We evaluate \sysName and its components through a series of benchmarks and a case-study. Our evaluations are all performed in unconstrained environments with our hardware system running in real-time. The benchmarks provide context of \sysName's AoA accuracy across various axes of variability: switching speed, number of antennas, and target location. 
Our case-study involves simultaneously estimating AoA of an iPhone, WiFi Access point (AP), and an Airpod in an unonstrained lobby environment. Our key results are compared with relevant prior work in Table~\ref{tab:comparison-table} and summarized below:

\begin{table}[]
\begin{tabular}{|l|l|l|l|}
\hline
Ref & \begin{tabular}[c]{@{}l@{}}Protocol\\ + devices\end{tabular} & \begin{tabular}[c]{@{}l@{}}Switched Rx\\ + architecture\end{tabular}                                                                              & \begin{tabular}[c]{@{}l@{}}AoA\\ Median\\ Error\end{tabular} \\ \hline
SWAN~\cite{wifiswitching-mobicom}     & \begin{tabular}[c]{@{}l@{}}WiFi+\\ single\end{tabular}       & \begin{tabular}[c]{@{}l@{}}9-element + \\ WPJ558 WiFI AP\end{tabular}     & 2.6$^\circ$                                                           \\ \hline
TYRLOC~\cite{tyrloc}     & \begin{tabular}[c]{@{}l@{}}BLE+\\ single\end{tabular}        & \begin{tabular}[c]{@{}l@{}}8-element +\\ PlutoSDR + FPGA\end{tabular}   & 2.9$^\circ$                                                           \\ \hline
\cite{ubicompbleswitch}     & \begin{tabular}[c]{@{}l@{}}BLE +\\ multi device\end{tabular} & \begin{tabular}[c]{@{}l@{}}8-element +\\ PlutoSDR\end{tabular}           & 2.37$^\circ$                                                         \\ \hline
\textbf{\sysName}    & \begin{tabular}[c]{@{}l@{}}\textbf{All} +\\ \textbf{multi device}\end{tabular} & \begin{tabular}[c]{@{}l@{}}8-element + \\ USRP B210 \\ + FPGA\end{tabular} &\textbf{ 1.4$^\circ$ }                                                         \\ \hline
\end{tabular}
\caption{Comparison of this work with prior switched antenna array AoA estimation techniques. \sysName is protocol agnostic.}
\label{tab:comparison-table}
\end{table}
\begin{enumerate}
    \item \sysName provides 90th percentile AoA accuracy 3.3$^\circ$, and median accuracy of 1.4$^\circ$, 1.7$\times$ better than baseline. This evaluation involved real BLE and WiFi signals.
    \item \sysName can separate devices operating simultaneously in a wide-band (30.72 MHz) and provide frequency, time context in addition to AoA.
    \item \sysName can estimate 8-antenna AoA for packets as small as 50 us. For smaller packets, performance degrades gracefully as it is equivalent to AoA estimation with lesser antennas.
\end{enumerate}

For experiments in the 2.4-2.6 GHz bands, we use \texttt{ARRAY2.4GHz} with 6.25 cm inter-antenna spacing ($\lambda$/2 for 2.4 GHz). For experiments in the 5 GHz band, we use \texttt{ARRAY6GHz} with 2.5 cm inter-antenna spacing ($\lambda$/2 for 6 GHz). Unless otherwise specified, we use an array with 9 antennas (8 switching antennas + 1 reference antenna) in the horizontal plane and the switching time between consecutive antennas is 6.25 $\mu$s. All ground truth AoA are measured manually using laser levels, protractors and range-finders. All measurements were made while observing a band 30.72 MHz wide using the USRP B210.

\subsection{Benchmarks}
\label{benchmarks}

\begin{figure}[!t]
    \includegraphics[width=1\columnwidth]{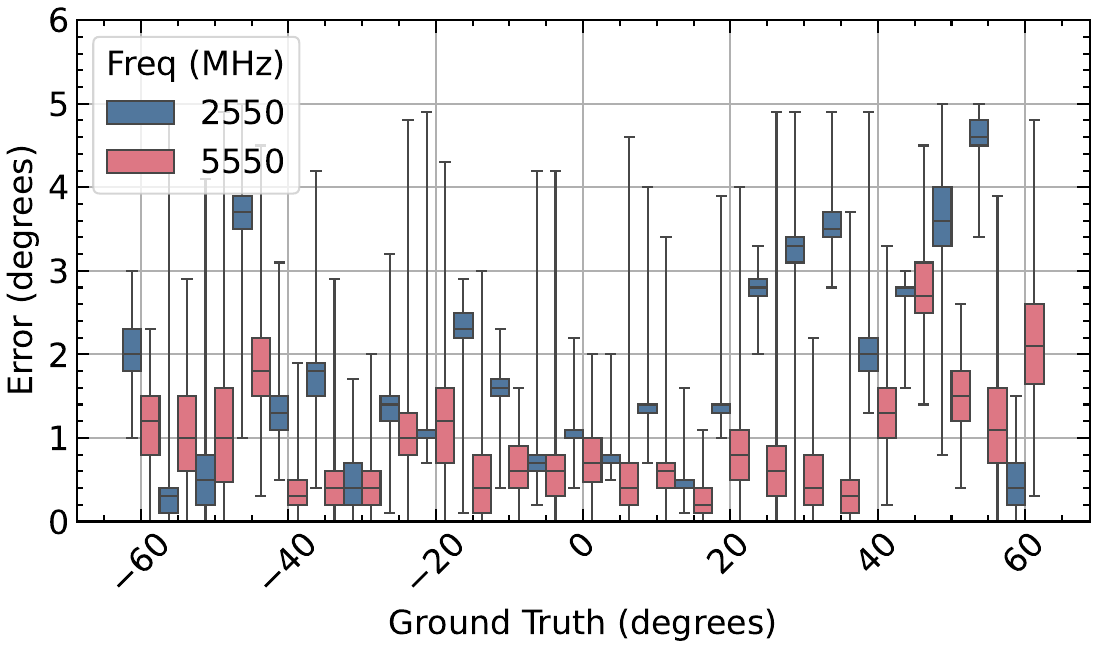}
    \caption{\small{AoA accuracy for two bands of interests as a function of ground truth AoA. Errors, especially as we scan wider angles frequency are due to uncalibrated radiation pattern of the antennas and the natural effect of scanning too wide on linear arrays}}
    \label{fig:aoa_accuracy}
\end{figure}

\noindent
\textbf{AoA accuracy across field of view}: The accuracy for \sysName was evaluated in an indoor setting by estimating the Azimuth AoA for a known transmitter. Evaluations were done for 2 different test signals. For the first evaluation, the transmitter was set to transmit OFDM packets at a center frequency of 2.55 GHz with bandwidth of 5 MHz. We use the horizontal \texttt{ARRAY2.4GHz} (6.5 cm) for this evaluation. The TX-RX separation was set to 2m with clear line of sight and both TX and RX at the same height. Each estimate requires only a single packet, and boxplots with estimates aggregated over 500 packets for every angle are shown in Fig~\ref{fig:aoa_accuracy}. At 2.55 GHz, the errors are all consistent with a spread of less than 3 degrees. The consistent errors, especially as we scan wider angles frequency are due to uncalibrated radiation pattern of the antennas at 2.55 GHz and the natural effect of scanning too wide on linear arrays~\cite{ubicompbleswitch}.

\noindent
\textbf{AoA accuracy in different frequency bands:} The second evaluation was done in a similar setting with the only difference being the center frequency of the packets being 5.5 GHz (we use \texttt{ARRAY6GHz} for this experiment), and the TX-RX separation being 1m (to account for the higher path loss for the same transmit power). Different AoA values in the range of -60 to 60 degrees were tested by rotating \sysName in different directions. Fig~\ref{fig:aoa_accuracy} shows a boxplot of errors at 5.5 GHz. Even in this frequency band, \sysName is able to estimate the AoA for a single transmitter with an accuracy of less than 2 degrees for all but two locations. While the errors increase for larger scan angles, they are much more consistent at 5.5 GHz.

\begin{figure}[!t]
    \includegraphics[width=1\columnwidth]{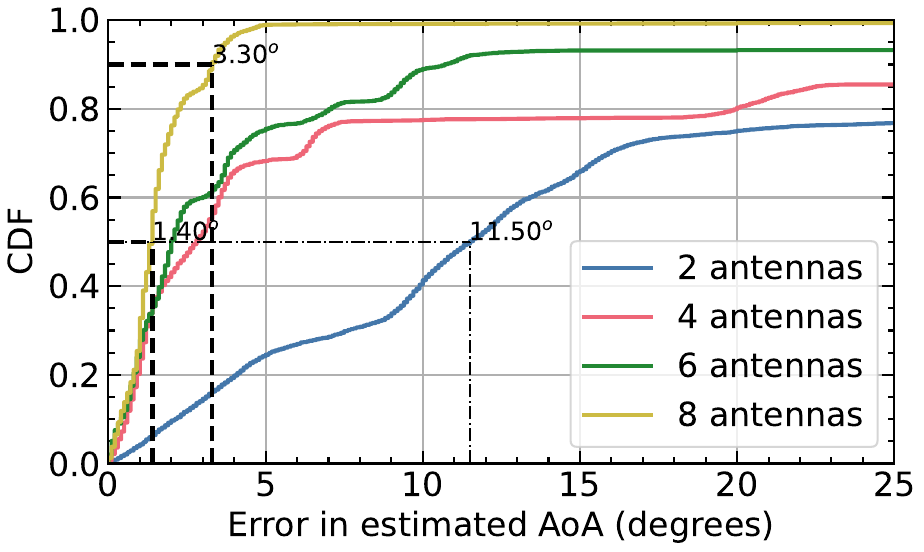}
    \caption{\small{CDF of estimated angle of arrival accuracy with varying number of antennas across all data collected at 2550 MHz. For the typical 8 antenna case, we get a median AoA accuracy of 1.4 degrees and a 90th percentile accuracy of 3.3 degrees}}
    \label{fig:cdf}
\end{figure}

\noindent
\textbf{AoA accuracy with number of antennas}: We plot the overall AoA accuracy from all our experiments 2.55 GHz to get a better picture of typical accuracy. In addition, we use saved artifacts from our real-time evaluation to compute performance of \sysName with 2, 4, and 6 antennas. Each CDF contains 6500 AoA estimate, each AoA estimate is performed on one packet, 500 estimates are made for each ground truth (13 sets spanning -60 to 60 degrees). The TX-RX separation was 2m with clear line of sight and both TX and RX at the same height. The TX was set so that both the ground truth Azimuth and Elevation AoA were 0 degrees. Figure~\ref{fig:cdf} shows the CDFs AoA for different number of antennas. For the typical 8 antenna case, we get a median AoA accuracy of 1.4 degrees and a 90th percentile accuracy of 3.3 degrees. Performance of \sysName with lower number of antennas gracefully degrades from 8 to 4 antennas, but there is close to 10 degree degradation in median error from 4 to 2 antennas. This is due to the fact that 2 antenna arrays have constrained scan angles and a very large beamwidth~\cite{rfidraw}.

\begin{figure}[t]
    \includegraphics[width=1\columnwidth]{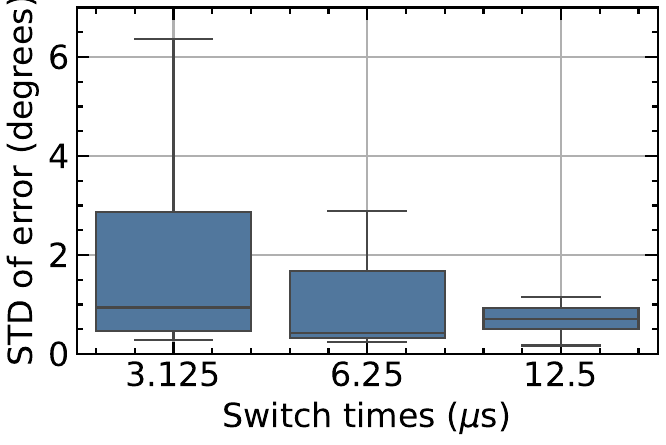}
    \caption{Impact of antenna Switch time on angle of arrival estimation accuracy. Smaller switch times cause to more distortion and less number of samples per switch interval, leading to larger error spread.}
    \label{fig:switchtimes}
\end{figure}

\noindent
\textbf{Effect of switching time}: So far, all the evaluations were done with a constant switching time of 6.25 $\mu$s which is much smaller than the time duration of most kinds of RF packets. Faster switching times enable data collection across more antennas. To get a better idea of the the impact of switch times, we tested multiple different switch times in experiments similar to the ones before. An OFDM signal center frequency 2.55 GHz and bandwidth 5 MHz was used as the transmitter and \sysName was used with \texttt{ARRAY2.4GHz} (6.5 cm). Multiple ground truth angles were evaluated and estimates from all the orientations were stored. Fig~\ref{fig:switchtimes} shows the standard deviation of estimates for different switch times. We see that the spread of AoA estimation error increases from $\approx$1 to $\approx$6 degrees as switching becomes faster. This is expected due to the lower number of samples per antenna and the additional distortion due to switching fast. Since \sysName uses tightly synchronized hardware and careful signal processing, such switching rates can be sustained. For context, the fastest switching array to our knowledge uses switching times that are $\ge$29 $\mu$s, close to 5x slower~\cite{ubicompbleswitch}.

\subsection{\textbf{Multi-device simultaneous AoA}}
\label{locating_indoors}
\begin{figure}[!t]
    \includegraphics[width=0.5\textwidth]{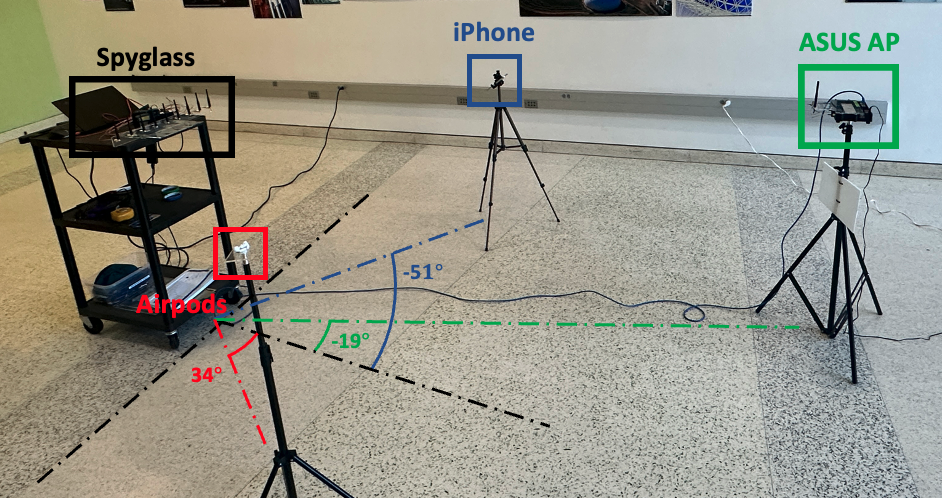}
    \caption{Multi-device AoA evaluation setup}
    \label{fig:multi_device_setup}
\end{figure}

The performance of \sysName in detecting real-world transmitters was also evaluated. Three devices - an iPhone transmitting iBeacon packets, a pair of Airpods Pro transmitting Bluetooth packets and an ASUS AP transmitting Wi-Fi packets on channel 1 were used for this evaluation. All three devices were placed randomly in a 5m $\times$ 4m environment and \sysName was used to estimate the angles at which these devices were placed. The setup along with random ground truth orientations can be seen in Fig \ref{fig:multi_device_setup}. 

\begin{figure*}
    \centering
    \begin{subfigure}{.49\textwidth}
        \includegraphics[width=\linewidth]{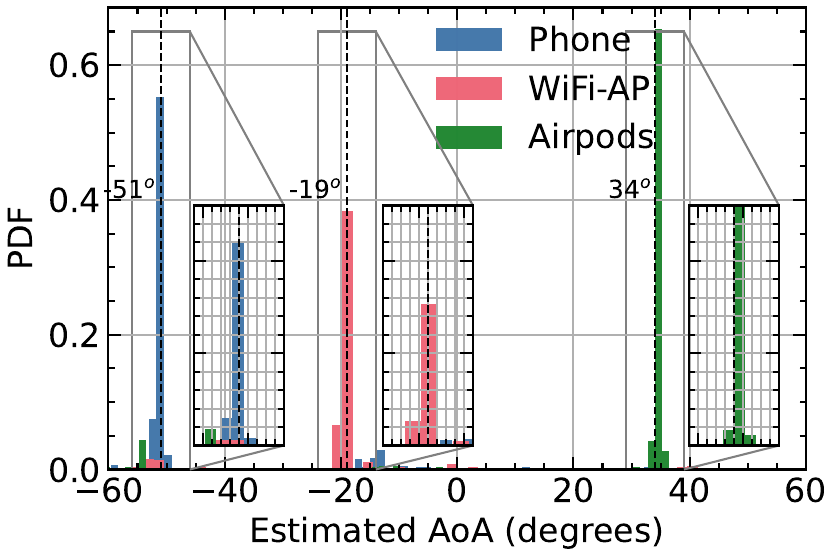}
        \caption{One device at a time}
        \label{fig:indv hist}
    \end{subfigure}
    \hfill
    \begin{subfigure}{.49\textwidth}
        \includegraphics[width=\linewidth]{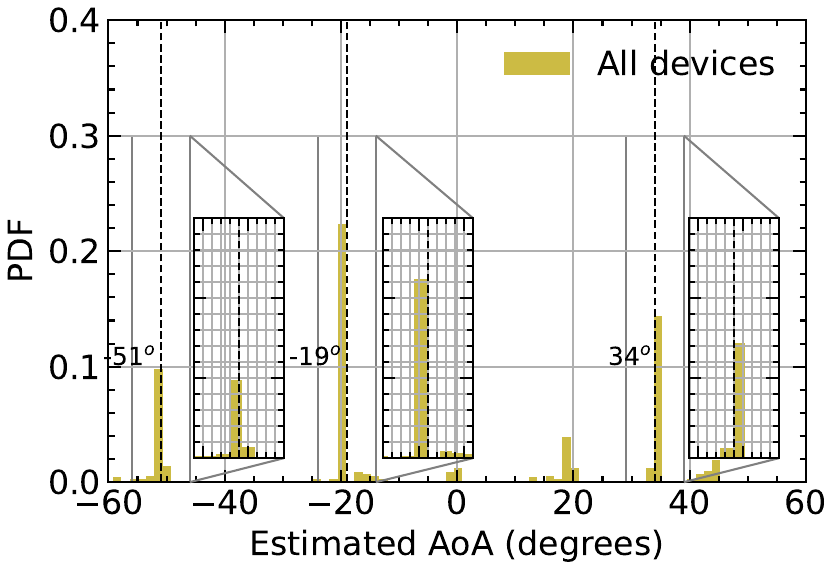}
        \caption{All three devices simultaneously}
        \label{fig:all devices}
    \end{subfigure}
    \caption{\small{Histogram of AoA estimation with three simultaneous devices: iPhone, airpods, and ASUS WiFi access point. Even when all three devices transmit simultaneously (b), \sysName can accurately estimate the AoA for each device and separate them in time-frequency and AoA.}}
    \label{fig:aoa_accuracy_multi_device}
\end{figure*}

 The AoA for each of the individual devices was first estimated by turning on one device and keeping all other off. To do this, a filter was used on the output energies of the energy detection algorithm which would only keep packets of certain center frequencies and bandwidths. This filter would be varies for different protocols being tested. Estimates were collected over 500 packets for each of the devices. The histograms of estimates along with the best estimate rounded to the closest degree are shown in Fig \ref{fig:indv hist}. It can be seen that \sysName is able to accurately detect the angles at which each of the devices are kept for approximately 70\% of detected packets. The other 30\% of estimates are likely incorrect due to effects of multi-path in indoor environments as well as detection of other devices not under test in the location.

Evaluations to get AoA information for all the devices above simultaneously were also done. All the devices were kept at the same locations with the same ground truth AoAs as before. The filter on energies was set to pass only Bluetooth packets and channel 1 Wi-Fi packets. AoA estimates were collected for 500 packets in total. The histogram of these estimates is shown in Fig. \ref{fig:all devices}. It can be seen from the figure that there are very strong peaks at three AoA values - -51 degrees, -19 degrees and 34 degrees (rounded to closest degree). These correspond correctly with the ground truth angles of the three devices under test. This shows that the \sysName system can accurately estimate AoA for multiple devices using multiple different protocols at the same time.

\subsection{Elevation AoA and multi-path effects}

\sysName's antenna arrays can be configured in any manner. We experimented with a 4x4 'L'-shaped array to simultaneously estimate azimuth ($\phi$) and elevation ($\theta$) AoA from a packet in a single shot. In all prior experiments, we only measured azimuth ($\phi$) AoA (horizontal plane). These evaluations were performed using the test OFDM waveform of 5 MHz bandwidth at 2.55 GHz. The evaluated points lie on the 8 corners of a 1 m x 1 m x 1 m cube centered 2.5 m from the center of the antenna array. The array has an inter-antenna spacing of 6.25 cm. We evaluate AoA for 500 packets at each point.

\begin{figure}[t]
    \includegraphics[width=\linewidth]{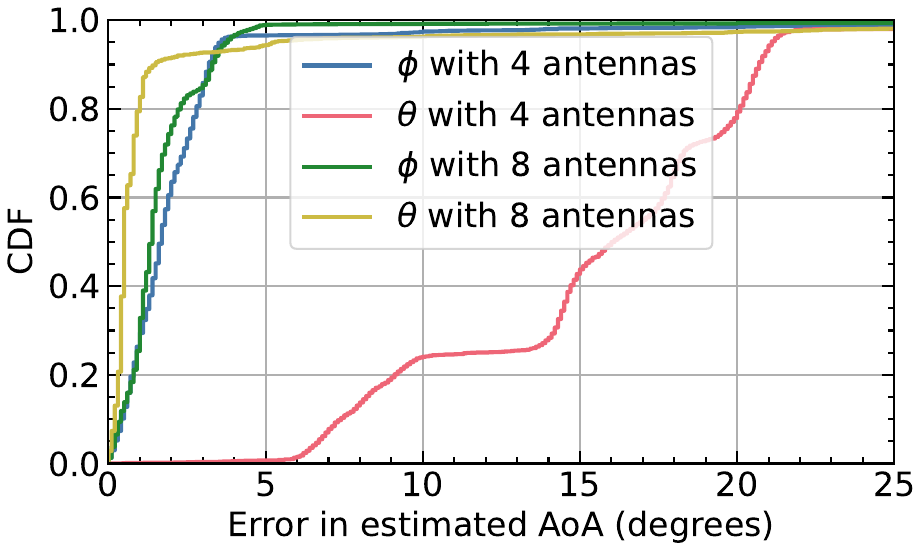}
    \caption{\small{CDF of AoA errors with different number of antennas used in the azimuth ($\phi$) and elevation ($\theta$) plane. 4 antennas AoA is erroneous for elevation due to multi-path from the floor and ceiling.}}
    \label{fig:azel_multipath}
\end{figure}

The CDF of AoA error is presented in Fig~\ref{fig:azel_multipath}. To our surprise, the accuracy of elevation ($\theta$) with 4 antennas is significantly worse with a median error of 17 degrees, as compared to the azimuth ($\phi$) median error of only 2 degrees! For comparison, we performed the same experiment once more with an 8 antenna vertical array, and saw that the CDF of errors is comparable to that of the 8 antenna horizontal array (also plotted in the same figure). We believe that the elevation measurements are more sensitive to multi-path effects due to strong reflections from the floor and ceiling in our experimental setup. Using 4 antennas in the vertical is not enough to separate the strong line of sight path from the clutter, but 8 antennas are enough.

\noindent
\textbf{Prevalence of multipath - a study:} Our study aimed to understand the prevalence of multipath in typical indoor environments. We utilized the DLOC CSI dataset~\cite{ayyalasomayajula2020deep} with azimuth AoA ground-truths, gathered in an 8m$\times$5m indoor area. A single source transmitted at 5.775 GHz with a bandwidth of 76.25 MHz, while data was collected using three commercial off-the-shelf (COTS) Access Points (APs), each equipped with a 4-antenna horizontal array.

Ground truth AoA was calculated using the dataset's provided device locations and orientations. This was compared against AoA estimates derived from the AoA-Time of Flight (ToF) profile, identifying the angle with the highest magnitude as the estimated AoA. The presence of a single peak in the AoA-ToF profile indicated a dominant signal path, while multiple peaks suggested the presence of multipath.

To determine potential AoA values, we applied a 50\% threshold to the normalized magnitude of the AoA-ToF profile. Regions above this threshold were analyzed, and their centroids were recorded as potential AoA values. The study also addressed the issue of estimation ambiguity due to the maximum scan angle of antenna arrays. We excluded measurements with ambiguous peaks to ensure accuracy, as these could not reliably indicate multipath presence.

Our findings showed that when the AoA was set to the angle of maximum magnitude on the AoA-ToF profile, around 60.56\% of data points had an AoA error of less than 10 degrees compared to the ground truth. However, 56.78\% of the remaining points had ground truth values within the ambiguous range, potentially leading to incorrect estimates. Notably, for data points with a ground truth angle lower than the maximum scan angle, the AoA-ToF profile mostly had fewer than three maxima, indicating a lesser presence of multipath.

 \begin{figure}[t]
    \includegraphics[width=\linewidth]{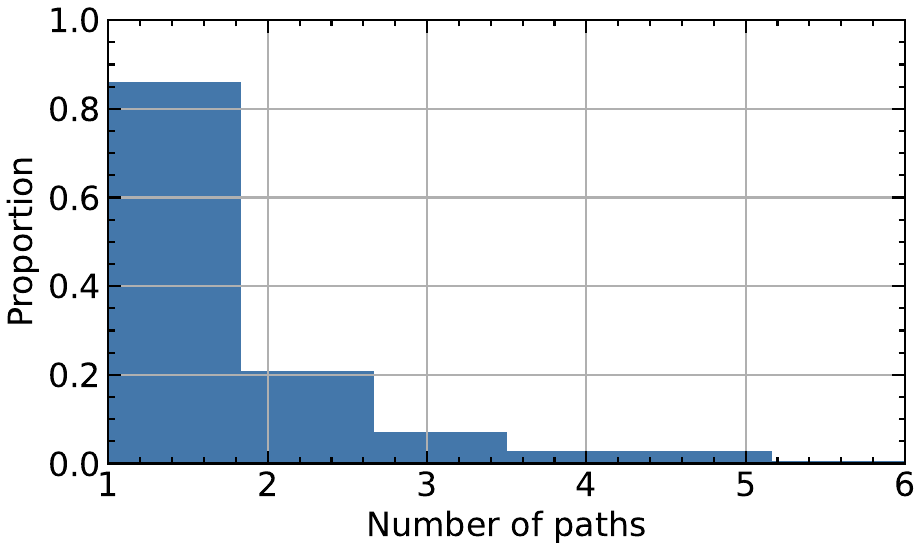}
    \caption{\small{Analysis of multipath in the Dloc dataset, showing that the LoS path is dominant in a large proportion of azimuth plane AoA estimation scenarios}}
    \label{fig:dloc_analysis}
\end{figure}

\subsection{Related Work}

\sysName is related to the following areas of research:

\textbf{Spectrum sensing}~\cite{ss1,sparsdr, bigband, bigband2} 
There has been a lot of interest in developing hardware platforms for spectrum sensing following the rise of spectrum sharing schemes like Citizen's Broadband Radio Spectrum (CBRS)~\cite{fcc_cbrs}. A number of them have focused on wideband spectrum observation~\cite{oneradio,mishali2011xampling_hw,ss1, bigband, bigband2}. BigBand~\cite{bigband} uses compressed sensing to sense occupancy of spectrum over a wide bandwidth. SweepSense~\cite{ss1} uses a sweeping LO to visit each band quickly and can perform rudimentary signal analysis. Some researchers have also focused on low-cost sensing hardware like ~\cite{lowcostkleber2017radiohound,ss1,lowcostluo2022wise}. All of these techniques focus on sensing spectrum occupancy and classifying it; some leaving further processing to the end user. \sysName can be used in tandem with hardware methods like ~\cite{ss1, lowcostluo2022wise} owing to the protocol agnostic nature of the \edalgo algorithm and the switching array. Directional spectrum sensing techniques ~\cite{directional1,directional2,directional3} leverage signal processing or directional antennas to sense direction in addition to aggregate spectrum occupancy, but cannot perform blind signal separation or extract AoA on each packet.

\textbf{Energy detection}: Algorithms to understand spectrum use based on energy and other features have been developed for many decades. Matched-filter and cyclostationary signal processing based methods use some prior knowledge of signals to design effective detectors~\cite{mf-cyclo-compare,gardner1992signal,cyclo-2}. More recently, deep-learning (DL) based methods have also been demonstrated~\cite{dl-ed1, dl-ed2, dl-ed3}. These DL methods require training and might not easily generalize to all spectrum environments. SparSDR~\cite{sparsdr} uses an invertible STFT and performs thresholding to compress I/Q samples at the SDR. In \edalgo, we use a similar invertible STFT, but a more robust noise-floor estimation algorithm combined with the ability to label and extract "boxes". In~\cite{bell2023searchlight}, the authors develop an optimal energy detection algorithm based on the polyphase filter-bank, aimed at detecting low-SNR anomalous signals. However,~\cite{bell2023searchlight} supports only a single channel, requires highly parallel processing, and is implemented on the GPU. In contrast, \edalgo uses simpler algorithms like the STFT to allow multi-channel energy detection on a laptop CPU.

\textbf{AoA Estimation for localization}: AoA estimation has been studied deeply for the purpose of localization~\cite{spotfi, arun2023xrloc, rfidraw, horus,pinpoint,pinit,wisee,arraytrack,chronos,xie2018md,xie2016xd,tonetrack,xiong2012towards,adib2013see,ubicarse,witrack}. Almost all of these techniques use simultaneous data from multiple-streams along with protocol specific channel estimation methods to extract clean relative channels. Using protocol specific channel estimates allows multi-axis AoA estimation using MUSIC and other super-resolution algorithms resilient to multi-path fading as in ~\cite{spotfi}. \sysName makes no assumptions of channel estimation, and performs blind AoA estimation, which cannot directly take advantage of wideband channel estimates. However, \sysName can make up for this deficiency by adding more antennas, which is low-overhead due to the switching array.

\textbf{Switched virtual arrays for localization}: Many low-cost localization methods have been developed using switched virtual arrays~\cite{ubicompbleswitch, wifiswitching-mobicom, tyrloc, gjengset2014phaser, adib2015capturing}. Some of these~\cite{ubicompbleswitch, wifiswitching-mobicom, tyrloc} use only one switched stream without a reference antenna, handling frequency offset issues algorithmically. All of these methods rely on specific prior knowledge of the target protocol.~\cite{tyrloc, wifiswitching-mobicom} switch antennas across multiple packets and are affected by collisions and inability to separate packets.~\cite{tyrloc} requires 21 packets to get one AoA estimate. In ~\cite{ubicompbleswitch}, the authors propose a sub-packet single channel switching method specific to the Bluetooth protocol.~\sysName can perform blind sub-packet switching over a wide-band without any assumption on signal protocol.

\section{Limitations and Discussion}

Developing and evaluating \sysName provided insights into how spectrum sensing and AoA estimation intersect, and into behavior of the wireless channels. Below, we list some of our learning and possible methods to extend \sysName in the future.

\noindent
\textbf{Synchronizing hardware with multi-channel SDRs is tractable:} Two channel SDRs are widely available in the market~\cite{limesdr, usrp, plutoplus, antsdr}, with many of them allowing synchronization signals to be exported as reference clocks or GPIOs. In addition to our work, multiple authors have shown promising results with synchronized hardware peripherals with SDRs~\cite{mimorph, crescendo, wintechsynced}. 

\noindent
\textbf{Handling multi-path, especially in elevation domain:} We noticed that AoA estimation in the elevation domain is challenging due to multipath reflections from the ceiling/floor and polarization effects. This effect can be handled by using more antennas to spatially separate multi-path. 

\noindent
\textbf{Augmenting \sysName with protocol specific information:} protocol specific channel estimators can be easily added to \sysName owing to the fact that the reference stream provides a clean path for estimation and even decoding. These channel estimates can be used with wide-band MUSIC~\cite{spotfi} to improve multi-path separation.

\noindent
\textbf{Extending to multi-device, any-protocol localization:} Deploying multiple \sysName systems in an environment can enable low-latency multi-device, any-protocol localization. This has significant practical applicability in spectrum management, digital twin, and in virtual spaces. We leave this for future work.

\section{Acknowledgments}

The authors are grateful for technical feedback from colleagues in WCSNG, UC San Diego. We acknowledge the support of NSF grants CNS-2213689 and OSI-2232481 for this research.


\end{document}